\documentclass[twocolumn]{aastex631}

\usepackage[T1]{fontenc}

\usepackage{graphicx}
\usepackage{hyperref}

\usepackage{txfonts}
\usepackage{multirow}

\usepackage{amsmath}
\usepackage{graphicx}
\usepackage{xcolor}
\usepackage{soul}

\begin{document}

\title{Equation of State for warm Neutron Star outer crusts}


\author[0000-0002-1518-5584]{David Barba-González}

\affiliation{Department of Fundamental Physics and IUFFyM, Universidad de Salamanca, Plaza de la Merced, s/n, E-37008 Salamanca, Spain}

\affiliation{Racah Institute of Physics, The Hebrew University, Jerusalem 9190401, Israel}
\email{david.barba-gonzalez@mail.huji.ac.il}

\author[0000-0002-0248-8260]{Conrado Albertus}
\affiliation{Department of Fundamental Physics and IUFFyM, Universidad de Salamanca, Plaza de la Merced, s/n, E-37008 Salamanca, Spain}
\email{albertus@usal.es}

\author[0000-0003-3355-3704]{M. Ángeles Pérez-García}
\affiliation{Department of Fundamental Physics and IUFFyM, Universidad de Salamanca, Plaza de la Merced, s/n, E-37008 Salamanca, Spain}
\email{mperezga@usal.es}


\begin{abstract}

We describe the equation of state (EoS) of a warm ion plasma as obtained by performing  microscopic many-body simulations using Molecular Dynamics computational techniques. Using the cold one-component plasma (OCP) composition in the Neutron Star (NS) outer crust assumed in \cite{Murarka_2022} with a representative heavy nucleus for each density, 
we refine previous calculations. We include electron screening and modeling of ions as finite-size Gaussian distributions in the interaction potential, together with an efficient Ewald energy summation procedure. From this, the EoS relation   $P(n_B,T)$ is obtained as a function of baryonic number density $n_B\in[7.48\times 10^{-10},2.09\times10^{-4}]$  $ \rm fm^{-3}$ and temperature $k_{B}T\in[1,5]$ MeV  in the NS outer crust conditions.  In order to improve the usability of our results we provide tabulated data values  along with a neural network parametrization available in the Zenodo repository, see \dataset[doi:10.5281/zenodo.15348712]{https://doi.org/10.5281/zenodo.15348712}. We find that even at moderate temperatures, thermal effects of ions are key in the higher density region closer to the inner crust, when described using a thermal effective parametrization based on the thermal adiabatic index $\Gamma_{th}$. We compare our results with other EoS in the literature performing a critical discussion. 
\end{abstract}



\section{Introduction}

\label{sec:intro} 

The microscopic description of warm low density matter of astrophysical interest is of uttermost importance when analysing  core-collapse supernovae (CCSNe), protoneutron star (PNS) evolution, and compact binary coalescence (CBC) events. The Equation of State (EoS) is the key ingredient to understand the internal dynamics in these events. Contrary to cold evolved objects, in these scenarios ultradense matter displays non-trivial thermal effects \citep{Raduta2021}. Analyses performed following computational simulations of low density warm matter show that plasma material expected in these events displays a thermodynamical behaviour capable of transitioning among gas-liquid-solid phases \citep{prc22,mnras24}  as it cools down \citep{Harutyunyan_2016}. Further, when thermal relaxation is put to a closer look it may display anomalous features \citep{anomalousmnras2025} differing from classical cooling laws. 

As a result of interaction at different scales, the full picture in these catastrophic events involves high gravitational/electromagnetic fields blended with complex micro-physical processes at strong and electroweak scales and offers the possibility to study them only indirectly from observations in multi-messenger scenarios. This has been only available since quite recently due to the latest technological progress, see \cite{2022A&A...665A..97R}. 

As an example, the recent detection of gravitational wave signals GW170817 \citep{Abbott2017} or GW190425 \citep{Abbott2020} originating from binary neutron star (BNS)  mergers detected by the LIGO and Virgo collaborations have opened up a new era. The confirmed Kilonova AT2017gfo \citep{Abbott2017} associated to the former GW170817 emission, single event of its kind, has allowed a pioneering test of our current understanding of its light curve and spectral features. To date, experimental features indicate they are sites of formation of elements in the lanthanide section due to \textit{r}-process decay of heavy nuclei \citep{Levan2024}. See \cite{https://doi.org/10.1002/andp.202200306} for a review on Kilonova ejecta and its microphysics dependences. 

In BNS events, the ejection of masses in the range $\sim (10^{-3}-10^{-2}) \ M_\odot$ takes place on a dynamical timescale during the violent disruption of the NSs \citep{Radice2018}. Matter is initially hot and neutron rich  with protons seeding nuclei that absorb the more abundant free neutrons, increasing their size. Due to the rapid neutron capture, it is believed  this mechanism leads to the formation of up to half of the heavy elements that we observe in nature. As mentioned in \cite{Perego2021} the prediction of the precise composition of the ejecta requires a detailed knowledge of the properties of exotic nuclei. Unfortunately this is out of reach at present and only with the advancement of rare isotope beam production over the last decade and improvement in experimental sensitivities, many of these rare-earth nuclides have now become increasingly accessible \citep{annurev:/content/journals/10.1146/annurev-nucl-121423-091501}. 

In these conditions, studied via strong-tidal field simulations, density and temperature are time evolving quantities.
Warm NS ionic crust matter, in the low density region below the crust-core transition appears in a phase that is no longer homogeneous, but forming cluster-like structures. It is already well known that competing long-range and short range interaction between nucleons drive frustrated configurations where matter shows non-fluid features, until it transitions to a material comprised of bound nuclei, see  \cite{BAYM1971225,horo2005,2021PhRvC.103e5810C,NewtonPhysRevC.105.025806} just to cite some examples. 
 In the intermediate evolutionary stages warm material composed of a majority of nuclei and to a lesser extent, nucleons are key to describe material properties. Neutrality is assured by electrons,  taken as relativistic and degenerate. Once the density, temperature and composition of this plasma are given, the entropy of the system is also known. In this scenario, the calculation of thermodynamical quantities, considering the reasonable assumption of ionized states, is performed assuming an idealized $\sim 1/r$ Coulomb potential interaction. 
However, in most of the calculations in the literature, dynamical correlations and screening effects has been typically overlooked, assuming that ions are mostly either an ideal gas or a cold static crystallized phase, depending on the scenario, in order to avoid the complexity of a microscopic treatment \citep{Yakovlev_2020,dehman2024A&A...687A.236D}. This results in a systematic overprediction of binding energies and, consequently, a bias in the low-density plasma EoS. 

In addition, a comprehensive description of the neutron star crust is often incomplete, as the composition is typically determined by minimizing a thermodynamic potential within the Wigner–Seitz cell approach \citep{PhysRevC.75.055806}.  Although this allows a reasonable reproduction of nuclear properties \citep{kazuo2008}, inherits the lack of correlations of dynamical cluster formation that can not be overlooked, this was discussed in \cite{horoPhysRevC.70.065806} in light of further impact on stellar opacities and, in general, on cooling properties.

Modern tabulated data provide an additional useful approach; see, for instance \citep{PhysRevC.73.035804} for the description of low density matter on cold scenarios in non accreting NSs. In the literature several crustal EoS aim to describe these outer layers using different treatments ranging from relativistic fields, Hartree–Fock–Bogoliubov calculations \cite{KREIM201363} or variational methods in the compressible liquid-drop model \citep{PhysRevC.76.065801}, just to cite some of them.  
For example, the energy density functional BSk24 \citep{eos_pcpbsk24} has been used to determine the ground-state composition and the EoS for a non-accreting NS.  Nuclear and electrostatic terms are introduced in a parametrized way to calculate both composition and thermodynamical properties \citep{eos_BL}, while other approaches 
try to unify the crust and core by employing a liquid-drop model in both regions 
of the EoS \citep{eos_GMSR}. Thermal effects are more cumbersome and effectively introduced in a minimal way  through so-called $\Gamma$--laws \citep{PhysRevD.98.043015} or ideal gas terms and Fermi gas corrections in the electron gas, see \cite{eos_HSDD2}, for example. These parameterizations, used in modern simulations, usually ignore ionic effects in the crust, ascribing thermal effects mostly to nucleonic degrees of freedom \citep{2023PhRvD.107b3016N}.

The microscopic technique used in this work, based on previous works \cite{prc22,mnras24,anomalousmnras2025}, allows for the description of  thermal effects along with dynamical correlations solved for a system of realistic ions in a warm plasma using screened Coulomb interactions. We perform computational Molecular Dynamics (MD) simulations of a warm system of finite size ions with Gaussian shapes superseding most previous point-like approaches, like    \citep{watanabe2003,horo2005,CaplanMulticomponent} and references therein.  For  ions with mass number beyond that of iron $A \gtrsim 60$, the relevant quantum scale is the thermal de Broglie wavelength $\lambda_{\rm dB} = \sqrt{2\pi \hbar^2/m_i k_B T}$ which is to be compared with the mean interionic spacing $a = \left( 3/4\pi n_i \right)^{1/3}$ also called Wigner-Seitz radius.  This approach is well grounded, since for typical warm dense conditions, $k_BT=1~\mathrm{MeV}$ and ions of mass number $A\in[60,100]$, the thermal de~Broglie wavelength $\lambda_{\rm dB}\sim [1.6,2.1]\times 10^{-5}~\text{\AA}$. For representative ionic number densities $n_i=10^{30}$–$10^{36}\,\mathrm{m^{-3}}$, the Wigner\textendash Seitz radius spans $0.62~\text{\AA}$ down to $0.0062~\text{\AA}$. Hence $\lambda_{\rm dB}/a \ll 1$ throughout, typically $\lambda_{\rm dB}/a \sim 10^{-5}-10^{-3}$, firmly placing the relevant ion dynamics in a semiclassical regime (we use $k_B=1$ units from now on).

This clear scale separation demonstrates that ionic motion can be reliably described within a MD framework, while electronic degrees of freedom remain quantum and provide the screening of the ion–ion interaction.

In this sense, a finite-size Gaussian representation of ions, in the spirit of Quantum Molecular Dynamics (QMD), is effectively recovered, although the wave-packet width is kept fixed in time. This treatment results in a more accurate calculation of the binding energy. As a result, a finite temperature EoS is obtained and analyzed, focusing on the actual thermal effective adiabatic index $\Gamma_{th}$, which is frequently used in current numerical relativity simulations.

{Contemporaneously with GW detection, another technique fueling a revolution not only on NS physics but science in general is artificial intelligence (AI). Within physics, one important capability is to proportionate easy-to-access numerical information outside of regions having been explored by modeling, see \cite{Jiao2024}. We take advantage of its power in this work, by using a Neural Network parametrization to characterize the outer crust EoS, and in this way microphysical properties of the ionic crust are easily and quickly accessible from our MD results.}

\section{Warm Neutron Star crust}
\label{sec:ocpcrust}

 In the region of interest for this work, which is the outer NS crust, nucleons (protons and neutrons) are mostly confined together forming nuclei along with a dilute neutron gas that may display superfluid features \citep{nucleisuperfluid}. In addition, charge neutrality involves the existence of an electron gas arising from those  electrons stripped off atoms in the vicinity of nuclei forming  a nearly degenerate sea around the positive nuclear charges, as $T \ll E_{F,e}$ . $E_{F,e}$ is the associated electron Fermi energy.

The treatment of electrons as a relativistic degenerate gas in a charge/mass $(Z,A)$ ionic system is usually sized by, first,  the relativistic parameter $x_{e,r}=p_{F,e} / m_e=\hbar\left(3 \pi^2 Z n_i\right)^{1 / 3}/m_e$ as the plasma is electrically charge neutral. Typically  $x_{e,r}\sim 1.008\left(\rho_6 Z/A\right)^{1 / 3}$   where $\rho_6=\rho / 10^6\,\rm g \,cm^{-3}$  and $p_{F,e}, m_e$ are the Fermi momentum and electron mass, respectively. Accordingly, the associated temperature $T_r \sim 5.930 \times 10^9 \mathrm{~K}$. Second, by the electron degeneracy parameter $x_{e,d}=E_{e,F}/T$. The electron-ion plasma is relativistic for $x_{e,r} \gg 1$ and $T \gg T_r$. Thus the electrons become relativistic beyond a few MeV thermal energy and baryonic number density $n_B \gtrsim 10^{-9}$ $\mathrm{fm}^{-3}$ while yet degenerate as $x_{e,d}\gg1$.

Inside the NS, in the cold and catalyzed regime, neutrons are expected to remain inside nuclei up until the neutron drip density $\rho_{\rm ND}\sim 10^{11} \rm g \ cm^{-3}$. Thus, at $T=0$ the outer crust mostly consists of  clusters or nuclei and their stripped off electrons around them. At earlier times, at higher temperatures, thermal excitations drive neutrons in a gas phase coexisting with nuclei at even lower densities \citep{pais2023A&A...679A.113P}. Note that as shown in \cite{Raduta2021} at temperatures of tens of MeV such as the ones expected in cooling proto-NS or BNS mergers, photo-production is also a source of pressure and energy density, and must be taken into account.

As already calculated for inner crust phases \citep{2023A&A...677A.174D}  the cluster population shows mass and composition spread according to the minimization of thermodynamical potentials involved according to  density and temperature. One class of such models concerns the one-component plasma (OCP) where one assumes that at each density there is only one possible ionic species.  The transition layers between densities have been recently modeled as having a finite thickness where more than one species can coexist due to e.g. accretion \citep{suleiman2024}, we focus here on the single ion OCP. This approximation has been shown to be valid even at low densities in the crust \citep{1984ApJ...285..294B}. Calculations within the Thomas-Fermi model predict the $(Z,A)$ at fixed density, typically overlooking further effects of neutron gas, as it is assumed to be negligible at very low temperature in the outer crust. The only remaining interactions are thus electromagnetic and finite-size effects on nuclei properties, such as volume or surface terms. Examples of these OCP, zero-temperature calculations for the outer crust are those of \cite{Murarka_2022} and \cite{pearson2018}  taylored to match properties of finite-size nuclei.

In our approach and with the aim of providing a sufficiently large density range for the warm NS crust EoS up to the outer/inner crust transition we have chosen a prescription based on the cold $T \sim 0$  composition. We must caution that our approach relies on the hypothesis that the composition stays unperturbed and the cluster dissolution does not take place up to temperatures about $\sim 5$ MeV. Recent works \citep{dehman2024A&A...687A.236D} based on microscopic Brueckner–Hartree–Fock calculations extended nuclear EOS calculations from cold to finite temperature for $\beta$-stable and alternatively, fixed proton fraction, in neutrino-free matter, focusing on the hot inner crust. They find that the crust-core transition density is temperature dependent in their OCP inherited approach. The critical data for constructing the outer crust EoS are the nuclear masses, sourced from the AME2012 evaluation \cite{Audi2012ChPhC..36....2A} or, if they are unknown, calculated using the Hartree-Fock-Bogoliubov (HFB) method with a prescribed functional. Binodal studies in \cite{menezesPhysRevC.60.024313} determined that in the context of non-linear Walecka models and also in supernova matter in nuclear statistical equilibrium \citep{ISHIZUKA2003517} the liquid-gas phase transition predicts cluster formation up to $T\sim 10$ MeV highly dependent on $Y_p$. This transition separates the uniform nuclear matter phase and a mixed phase in which the nucleons either populate impurities in the form of clusters, or sit in a gas in between them. These fractions of baryons clustered in the ions belonging to OCP, $X_i$, and the nucleon (neutron) gas $X_g\equiv X_n$ depend on both density and temperature. According to this prescription, the baryonic number density, in baryons per $\rm fm^3$, is given by 
\begin{equation}
    n_B=n_{B,i}+n_{B,n},
\end{equation}
where we set the fraction of ions and neutrons such that $X_{i}+X_n=1$ with  $n_{B,i}=X_in_B$ and $n_{B,n}=X_n n_B$. In addition there will be an electron degenerate sea in order to maintain charge neutrality with a fraction $n_e=Y_e n_B$. This corresponds to the total lepton density assuming electrons to be the only lepton source of the crust. We will further assume in our picture that the ion fluid and the neutron gas will not interact at hadronic scales at the outer crust lower densities we focus on in this work but the neutron gas is indeed present \citep{Grams2024}. Thus partial pressures can be calculated at their corresponding densities at a given crust depth, and added together to obtain the total baryonic pressure.

In terms of densities, as described in \cite{Grams2024}  the baryon density should contain the cluster (cl) and neutron gas (g) contributions, as $n_B=u n_{\mathrm{cl}}+(1-u) n_g$
where $u$ is the volume fraction or the ratio between the cluster volume over the Wigner-Seitz volume. There the energy density in  the dilute neutron matter is defined as $\epsilon_{\mathrm{tot}} \equiv \epsilon_{\text {mass }}+\epsilon_{\text {int }}$, where $\epsilon_{\text {mass }}$ is the rest mass energy density $\sim m^*_nc^2$ and the internal energy density $\epsilon_{\text {int }}=\xi E_{FG}+\epsilon_{\text {corr }}$ with $E_{FG}$ the Fermi gas energy and a factor $\xi\in [0.8,1]$ and the correlation part is approximated by $
\epsilon_{\text {corr }}=-\frac{1}{2} N_{0 n}\left(\Delta_n\left(n_n\right)\right)^2,
$ where appears the density of states $N_{0 n}=m_n^* k_{F_n} /\left(\pi^2 \hbar^2\right)$, the in-medium mass $m_n^*$, the Fermi momentum $k_{F_n}$ defined as $n_n=k_{F_n}^3 /\left(3 \pi^2\right)$, and a field $\Delta_n\left(n_n\right)$ which is adjusted to the 'ab initio' predictions for dilute neutron matter and amounts to $\sim 0.1E_{FG}$ for $n_n \lesssim 10^{-4}$ $\rm fm^{-3}$. The solution to stability equations yields a gas fraction that for the densities relevant for this work is $\lesssim 10 \%$ at $T=0$. At finite temperature, models yield diverse results. 
Nuclear Statistical Equilibrium (NSE) calculations \citep{eos_HSDD2} indicate that heavy nuclei account for a significant fraction of the baryon density from $\sim 10^{-7}\,\mathrm{fm^{-3}}$ up to neutron drip at $T=1\,\mathrm{MeV}$. Complementarily, relativistic mean field calculations (see Fig.~8 in \cite{ISHIZUKA2003517}) predict gas fractions of up to $\sim 10\%$ at densities $\gtrsim 10^{-9}\,\mathrm{fm^{-3}}$ under thermodynamic conditions relevant to this work. Ions thus can exert a relevant effect on thermal properties as long as matter is in the nuclear mixed phase, and will be especially relevant at densities close to the drip, as we will show below. 


The effective interaction potential is taken under a generalized Yukawa fashion. For a point-like source it adopts the familiar form $
V(r) \sim (Z_{\rm eff}^2 e^2/r) \, e^{-\kappa r}$ where \(Z_{\rm eff}\) is the effective ionic charge and \(\kappa\) is the inverse screening length. The latter is determined either by classical Debye theory for non-degenerate electrons or by finite-temperature Thomas–Fermi theory when electron degeneracy becomes important. The physics of this point-like system is primarily governed by two dimensionless parameters: the Coulomb coupling parameter, $\Gamma_C = Z_{\rm eff}^2 e^2/a k_B T$,
and the screening parameter \(\kappa a\). Together, these quantities determine whether the system behaves as a weakly correlated plasma, a strongly coupled liquid, or approaches a Coulomb solid. Very recently the importance of extended sources i.e. charge distribution spread have proved to be highly non-trivial, prescribing the need of an additional parameter, regarding charge spread or Gaussian finite-width \citep{prc22}.

It is important to mention that in QMD electrons are treated explicitly within density functional theory (DFT) or related ab-initio frameworks, while the ions move classically under forces obtained self-consistently from the quantum electron distribution. Such simulations naturally capture effects such as partial ionization, non-linear screening, and electronic transport properties. However, they are computationally far more demanding and limited to smaller system sizes and shorter timescales. For intermediate mass ions with \(A \gtrsim 60\) in warm dense plasmas, where \(\lambda_{\rm dB} \ll a\), the MD approach with screened interactions provides a physically justified and efficient description of ionic correlations, while the quantum aspects of the electrons are incorporated through the screening length \(\kappa\). Thus, MD with Yukawa interactions represents a robust and well-founded method for exploring strongly coupled ion plasmas in the warm dense matter regime.
Under these conditions, MD with Yukawa interactions reasonably describes  equivalent ionic correlations 
and transport properties that a QMD  treatment would yield. 
QMD becomes necessary only if one requires explicit electronic properties  or considers much lighter ions for which $\lambda_{\rm dB}$ does not fulfill being small compared 
to $a$. Thus, for the charge finite-spread ionic dynamics of warm neutron-star crust matter, Yukawa-based MD  provides a robust and efficient framework.


For the aforementioned electrons we treat them as a degenerate and ultrarelativistic Fermi gas screening the Coulomb interaction and adding an electron EoS $P_{FG,e}(\epsilon_{FG,e})$, whose energy density is given at $T=0$ as 

\begin{equation}
\epsilon_{FG,e}^{T=0} = \frac{1}{8\pi^2\left(\hbar c\right)^3}\left(\mu_{e} p_{F,e}\left(2\mu_{e}^2-m_{e}^2\right)-m_{e}^4 \mathrm{log}\left(\frac{\mu_{e}+p_{F,e}}{m_{e}}\right)\right).
\label{eq:epselec}
\end{equation}

From this one can calculate the pressure
\begin{equation}
    P_{FG,e}^{T=0} = -\epsilon_{FG,e}^{T=0}+\mu_{e}n_{e},
\end{equation}
where the chemical potential is given by $\mu_{e}=(m_{e}^2+\hbar k_{F,e})^{1/2}$ and $k_{F,e}= (3\pi^2n_{e})^{1/3}$ is the electron Fermi wave vector. At finite temperature this is no longer valid, although assuming $T<\mu_{e}$ one can apply the Sommerfeld expansion of the Fermi-Dirac distribution to obtain the low temperature expansion of the pressure density

\begin{equation}
    P_{GF,e}=P_{FG,e}^{T=0}\left[1+\left(\frac{5}{12\pi^2}\right)\left(\frac{k_BT}{\mu_{e}}\right)^2\right],
    \label{eq:Pelec}
\end{equation}
with an equivalent version for energy density.

A photon gas may appear at the temperatures expected at proto-NS or Binary Neutron Star mergers, so that the full pressure in this contribution would be 

\begin{equation}
    P = P_i+P_{FG,e}+P_{FG,n}+P_{ph},
    \label{eq:Ptot}
\end{equation}
where $P_i$ is the pressure associated with the combined  electromagnetic interaction between ions and ions-electrons, and $P_{FG,e}$, $P_{FG,n}$ and $P_{ph}$ are the electron, neutron and photon pressure, respectively. We can write a similar expression for the energy density as

\begin{equation}
    \epsilon=\epsilon_i+\epsilon_{FG,e}+\epsilon_{FG,n}+\epsilon_{ph},
\end{equation}
and we would obtain the EoS.

One of the main advantages of the methods used in this work is the access to finite temperature, provided that the degrees of freedom remain meaningful along with thermal effects. With the aim of comparing our results with those at zero temperature in the literature, we make use of the so-called $\Gamma$-factor, also known as the thermal adiabatic index. It is widely used to produce extensions of cold equations of state to finite temperature in a simple way.  We define the thermal energy density as 

\begin{equation}
    \epsilon_{th}=\epsilon-\epsilon(T=0),
    \label{eq:eth}
\end{equation}
and its pressure counterpart

\begin{equation}
    P_{th}=P-P(T=0).
    \label{eq:pth}
\end{equation}
In this manner the total pressure can be written as 

\begin{equation}
    P=P(T=0)+(\Gamma_{th}-1)\epsilon_{th},
\end{equation}
where the thermal index is defined as

\begin{equation}
    \Gamma_{th}=1+\frac{P_{th}}{\epsilon_{th}}.
    \label{gamamth}
\end{equation}

Hydrodynamical simulations usually take $\Gamma_{th}\in[1.5,2]$, however,  for  several EoS, predicted values lie outside of that range, including the relativistic Fermi gas in Eqs. \eqref{eq:Pelec} and \eqref{eq:epselec} for which  $\Gamma_{th}=4/3$ \citep{Raduta2021}. Another useful limit applicable in the crust is that of a dilute gas with tiny interactions between particles, the ideal gas predicts $\Gamma_{th}=5/3$.  

\subsection{Ionic contribution to the EoS}
For the ionic sector, and in order to capture finite temperature effects in a realistic manner, we focus on the widely-known technique of Molecular Dynamics. By selecting adequate ranges of density and temperature, we can approximate the crust as a one-component plasma of ions, embedded in a fully degenerate and relativistic electron gas. In our previous works \citep{prc22,mnras24}, we have studied crystallization and finite-temperature effects in reduced composition samples. We now generalize it to the full NS crust with composition data as a function of baryonic density taken from \cite{Murarka_2022}. Both the energy and pressure in the ionic sector are produced by electrostatic forces between ions and between ions and electrons. The electron sea where the ions are immersed is known to be polarized by their presence. It is not uniform, and the degenerate charge density is clustered around the ionic positions $\vec{r}_i$ as $e^{-|\vec{r}-\vec{r}_i|/\lambda}/4\pi \lambda^2$. In this work, electron responsiveness is codified in the dielectric function  valid for small particle momenta and small screening parameter $k_{TF}=\left(k_{F,e}+m_e^2\right)^{-1/4}\sqrt{\pi/4\alpha k_{F,e}}$ being the Thomas-Fermi wavevector. This means that the interaction between ions is effectively screened by electrons. 

In our setting, apart from this electron-induced screening, we also include finite-size effects for ions within an Ewald summation paradigm, which we have shown to have an effect in both crystallization and virial, high temperature properties. We refer the reader to our previous work, e.g. \cite{anomalousmnras2025} for the full expressions used to evolve the ionic samples, and reproduce here the energy and pressure that enter into the EoS. They are fundamentally based in the electrostatic potential that a Gaussian charge density with charge $Z_i$ and inverse square width $a_i$ exerts on its surroundings. This is

\begin{multline}
    \phi_{Z_i,a_i}\left(\vec{r}\right)=\frac{Z_i}{2|\vec{r}-\vec{r_i}|} e^{\frac{1}{4 a_i \lambda^2}}\left[e^{-\frac{|\vec{r}-\vec{r_i}|}{\lambda}}\mathrm{erfc}\left(\frac{1}{2\sqrt{a_i}\lambda} - \right.\right.\\ 
    \left.\left. \sqrt{a_i}|\vec{r}-\vec{r_i}|\right)  -e^{\frac{|\vec{r}-\vec{r_i}|}{\lambda}}\mathrm{erfc}\left(\frac{1}{2\sqrt{a_i}\lambda}+\sqrt{a_i}|\vec{r}-\vec{r_i}|\right) \right].
\end{multline}

Here we are taking the screening length $\lambda=k_{TF}^{-1}$. Thus, the total energy includes kinetic, potential and self-interaction contributions 

\begin{equation}
    U = U_{kin}+U_{\mathrm{short-range}}+U_{\rm long-range}-U_{\rm self}.
    \label{eq:Utot}
\end{equation}

The last three terms of Eq.~\eqref{eq:Utot}, concerning the electrostatic potential energy, are due to the Ewald summation separation between short- and long-range parts. The short-range part of the energy can be shown to be for N ions

\begin{multline}
    U_{\mathrm{short-range}}=\frac{1}{2}\sum_{i=1}^{N}\sum_{j\neq i=1}^{N} 2 Z_j\left(\frac{a_j}{\pi}\right)^{\frac{1}{2}} \frac{e^{-a_j r_{ij}^2}}{r_{ij}} \times\\ \int_{0}^{\infty} r' \phi_{\rm{short-range},i}\left(r'\right)e^{-a r'^{2}}\mathrm{sinh} \left(2a_jr_{ij}r'\right)dr' \\
    -\frac{2\pi}{V} \sum_{i=1}^{N}\sum_{j=1}^{N} Z_j \int_0^{\infty} r'^{2}\phi_{\rm{short-range}, \mathrm{i}} \left(r'\right)dr',
    \label{Ushort}
\end{multline}
where $\phi_{\mathrm{short-range}}=\phi_{Z_i,a_i}-\phi_{Z_i,\alpha_{\rm Ewald}}$. The parameter $\alpha_{\mathrm{Ewald}}$ does not affect the physics of the sample, and its choice is explained below. On the other hand the long-range part, summed over reciprocal space vectors $\vec{k}$, reads

\begin{align}
    U_{\mathrm{long-range}}= \frac{1}{2}\sum_{i,j=1}^{N} \sum_{\vec{k}\neq 0} \frac{4\pi Z_i Z_j}{V \left(k^2+\frac{1}{\lambda^2}\right)} \times 
    e^{\frac{-k^2}{4}\left(\frac{1}{a_i}+\frac{1}{\alpha_{\mathrm{Ewald}}}\right)}e^{i \vec{k}\left(\vec{r_i}-\vec{r_j}\right)}.
\label{Ulong}
 \end{align}

Finally, one needs to substract the energy added by spurious charge densities associated with the Ewald technique

\begin{equation}
    U_{\mathrm{self}}=2\pi\sum_{i=1}^{N} \left(\frac{a_i}{\pi}\right)^{\frac{3}{2}}Z_i \int_0^{\infty} r'^{2} \phi_{Z_i,\alpha_{\mathrm{Ewald}}}\left(r'\right) e^{-a_i r'^{2}} dr'.
\end{equation}

The forces the ions feel during the simulation are obtained by taking the gradient of the above expressions. Using such forces that particles exert on one another we can calculate the total stress tensor of the sample as

\begin{equation}
    \Pi_{\alpha \beta}^{\mathrm{tot}}=\frac{1}{L^3} \sum_{i=1}^{N_I} m_I \dot{\vec{r}}_{i \alpha} \dot{\vec{r}}_{i \beta}+\Pi_{\alpha \beta},
    \label{eq:Pitot}
\end{equation}
where $\alpha,\beta=1,2,3$.
Here, $\Pi_{\alpha\beta}=\Pi_{\alpha \beta}^{\mathrm{real}}+\Pi_{\alpha \beta}^{\mathrm{recip}}+\Pi_{\alpha\beta}^{\mathrm{volume}}$ due to the split in real and Fourier space mentioned above. From Eq. \eqref{eq:Pitot} we can obtain the total ionic pressure as $P_{i}=\frac{1}{3}\mathrm{Tr}\Pi_{\alpha\beta}$. We use the short-range, real forces to calculate 

\begin{equation}
\Pi_{\alpha \beta}^{\text {real }}=\frac{1}{2 V} \sum_{i,{j\ne i}=1}^{N}\left(F_{i j, \alpha}^{\text {real }} r_{i j, \beta}+F_{i j, \beta}^{\mathrm{real}} r_{i j, \alpha}\right),
\end{equation}
while the reciprocal, long-range sector is given by 

\begin{equation}
\begin{aligned}
\Pi^{\text {recip }}_{\alpha \beta}&=  \frac{1}{2}\sum_{i,j}^{N}\sum_{\vec{k}\neq0} \frac{4\pi Z_i Z_j}{V^2\left(k^{2}+\frac{1}{\lambda^2}\right)}e^{\frac{-k^{2}}{4}\left(\frac{1}{a_i}+\frac{1}{\alpha_{\rm Ewald}}\right)}e^{i\vec{k}\cdot\left(\vec{r}_i-\vec{r}_j\right)} \times \\
&\left\{\delta_{\alpha \beta}-2 \left[ \frac{1}{4}\left(\frac{1}{a_i}+\frac{1}{\alpha_{\rm Ewald}}\right)+\frac{1}{\left(k^2+\frac{1}{\lambda^2}\right)}\right]k_{\alpha} k_{\beta} \right\}
\end{aligned}
\label{reciprocal}
\end{equation}
and the final part, associated with the last term in Eq. \eqref{Ushort}, reads

\begin{equation}
    \Pi_{\alpha\beta}^{\mathrm{volume}}=-\frac{2\pi\delta_{\alpha\beta}}{V^2}\sum_{i=1}^{N}\sum_{j=1}^{N} Z_j \int_0^{\infty} r'^{2}\phi_{\rm{short-range}, \mathrm{i}} \left(r'\right)dr'.
\end{equation}

In order to evolve the sample, we integrate the ions' equations of motion under the above potential energy expressions. We use the Velocity Verlet algorithm, whose timestep $\Delta t$ we choose so that $c\Delta t /L_{\rm box}=0.05$ for each density in the crust. The quantity $c\Delta t$ represents only an upper kinematic bound on the displacement during one timestep. The actual ion displacement is T dependent $\Delta r=\langle v_i\rangle\Delta t$, with $\langle v_i\rangle$ obtained from the Maxwell--Boltzmann distribution. For the range of temperatures considered up to 5 MeV, this gives $\Delta r<D_i$, i.e. less than one nuclear (ion) diameter per timestep. The parameter $\alpha_{\rm Ewald}$ serves as a way of adjusting, within the Ewald summation, how much weight is given to the short- or long-range sector. In the simulations presented in this work, it is adjusted for each density and composition such that the short-range part can be safely cut off at the first layer of periodic simulation boxes, and thus we can use the minimum-image convention in the simulation box \citep{2013JChPh.139x4108H}. 

Simulations are run for at least $5\times 10^4$ time steps to assure thermalization, for $N=500$ ions at each density and temperature, in a parallelized OpenMP setup to ensure computational efficiency. Ewald summation ensures accurate results for moderate system sizes.  As seen in the Table \eqref{tab:md_outer_crust} in the appendix,  we span a range of densities comprising the whole outer crust up to $n_{B,i}=2.0938\times 10^{-4}\ \rm fm^{-3}$, corresponding to the neutron drip density $\sim 4\times 10^{11} \rm g \ cm^{-3}$, for different temperatures from $k_BT=1,2,3.5,5$ MeV. We will show below how the density and temperature range lie in the uncertainty region around the spinodal curve marking the nuclear phase transition to a liquid phase. We note that in this work the MD configurations correspond to warm liquid phases, our focus being on the thermodynamical properties of the system before forming a crystal lattice. The thermodynamical quantities, i.e. kinetic and electrostatic energy and pressure are extracted by averaging the quantities obtained from Eqs. \eqref{eq:Utot} and \eqref{eq:Pitot} after thermalization, which is forced upon the system by rescaling the velocity of the ions to drive them to the desired value for $T$ while maintaining the correct distribution for the velocities. This mechanism proves efficient at thermal and energy control, being at the instantaneous uncertainty percentage  $\lesssim 5\%$ for all simulations in this contribution.

\section{Screened OCP and Neural networks}

\begin{figure}
    \centering
    \includegraphics[width=\linewidth]{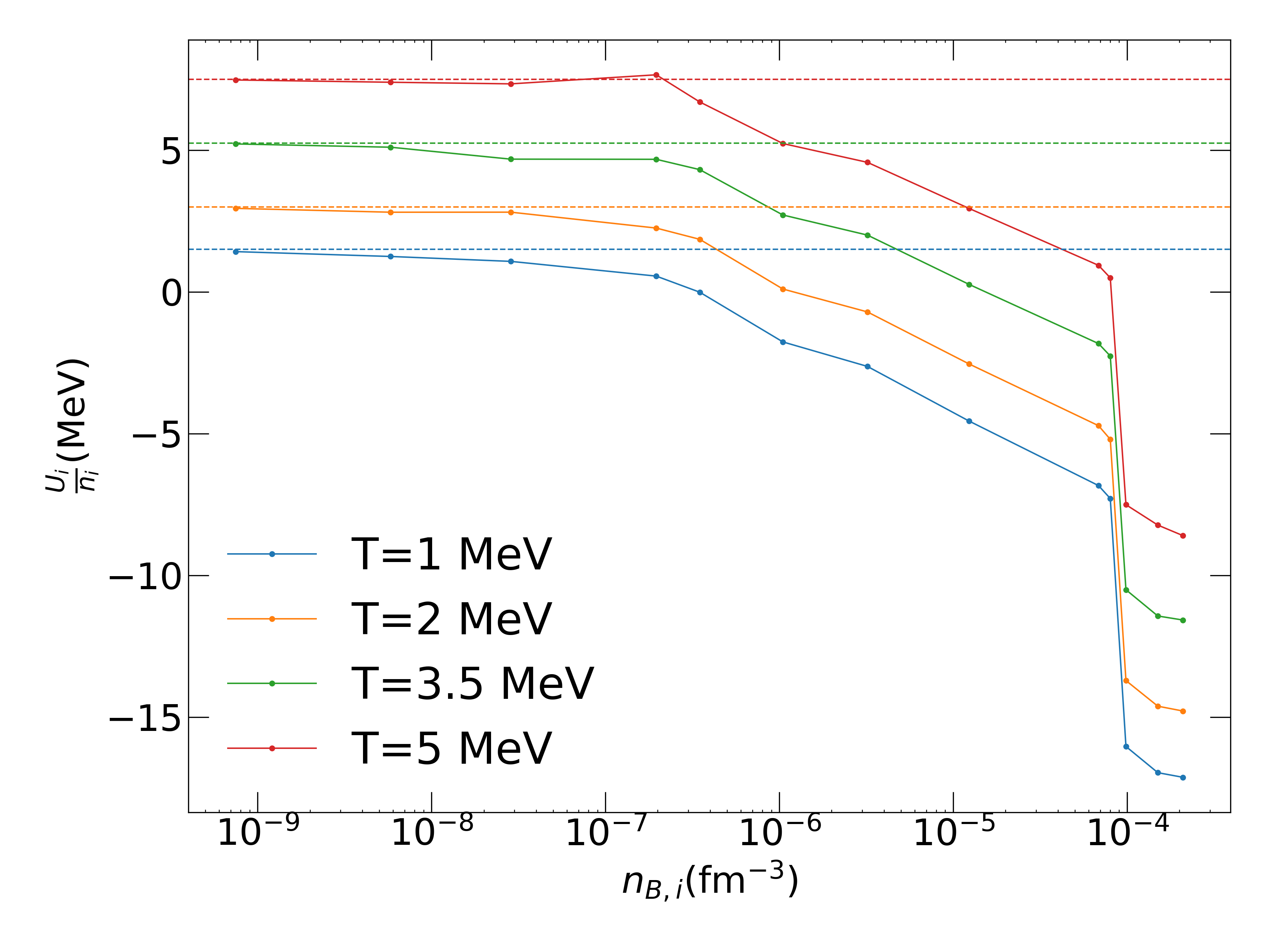}
    \caption{ Energy (from Eq.~\eqref{eq:Utot}) as a function of baryonic density for ions. Solid lines correspond to the results of our microscopical simulations, while the ideal gas limit for each given temperature is shown as a dashed line, to showcase the system reaching it at sufficiently small densities. }
    \label{fig:energy_idealgas}
\end{figure}

\begin{figure}
    \centering
    \includegraphics[width=\linewidth]{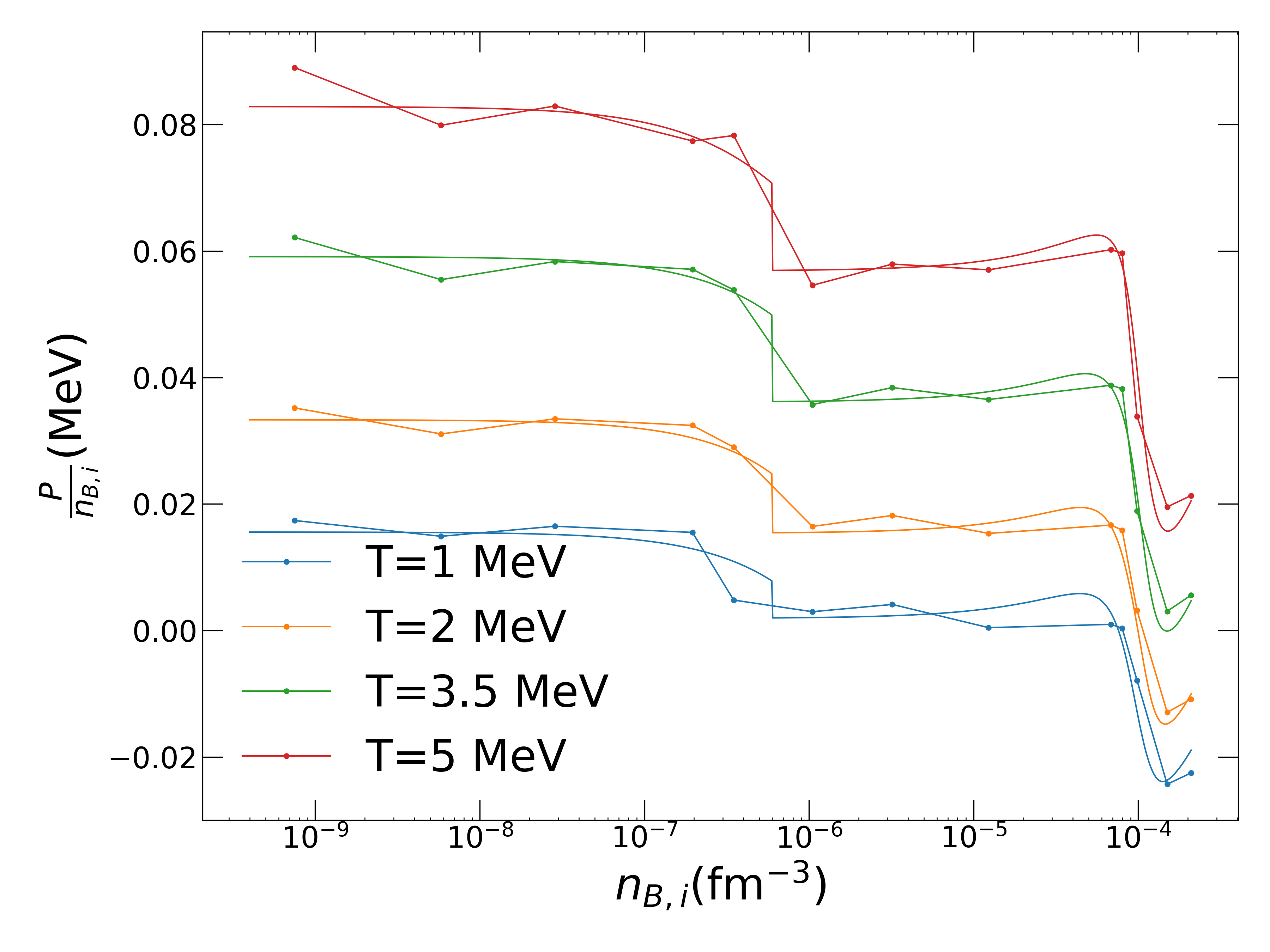}
    \caption{Points depict MD results for pressure as a function of baryonic density within ions for different temperatures. Smooth, solid lines are predictions from our neural network prescription, with goodness of fit $R^2>0.989718$ when compared to the simulation's results.}
    \label{fig:pressure_neuralnetwork}
\end{figure}

The composition of the crust is usually inferred by working in the context of a OCP approximation. This assumes for each density in the crust a characteristic ion (Z, A) as the sole contributor to the composition of the material. As mentioned, we assume the composition from \cite{Murarka_2022} for our simulations. Thus, at each value of the density we take the sample to be of a uniform composition with the same ion species for the full temperature range.

We show in Fig.~\eqref{fig:energy_idealgas} the energy density as a function of baryonic density in the crust for all available runs, see Table~\eqref{tab:md_outer_crust}. It can be seen that a departure from ideal gas behavior occurs at smaller densities as $T$ grows. The electrostatic energy density even becomes negative at sufficiently high densities, showcasing a transition to an ionic bulk bound state.

With the aim of providing the community with convenient access to the data presented in this paper, we have developed a Neural Network parametrization for the pressure as a function of baryonic density for the outer crust. As from Fig. \eqref{fig:energy_idealgas} there are two clear sections in density, we have separated the Neural Network in two different sets of parameters trained over the MD data. In particular, for the low-density region we use a triple hidden layer composed of three neurons, whereas in the high density two layers with two and three neurons each were enough to reproduce the data with high fidelity. The activation functions chosen were hyperbolic tangents. Both the MD results and the Neural Network that has been trained on them are fully available in a Zenodo repository \footnote{ https://zenodo.org/records/15348712}.

These two parametrizations, for high- and low- density, are presented together in Fig. \eqref{fig:pressure_neuralnetwork}, where the transition point is located at approximately $n_{B,i} \sim 6 \times 10^{-7}~\text{fm}^{-3}$. The plot also includes the original simulation data used for training. In both low- and high-density sectors of the outer crust, the agreement between the Neural Network predictions and the simulation data is excellent, with a coefficient of determination $R^2 > 0.989$, confirming the robustness of the fit across the full density range considered. Both sudden changes in pressure at $\sim 10^{-6}$ and $\sim 10^{-4} \ \rm fm^{-3}$ are due to changes in composition appearing in \cite{Murarka_2022} as the proton content inside the nucleus grows with baryonic density as we probe deeper in the outer crust.

In Fig. \eqref{fig:colorfig}, we present the total EoS constructed by combining the Neural Network parametrization of the electrostatic (ionic) sector with the pressure and energy density from the ultrarelativistic electron gas. The Molecular Dynamics data points are displayed as colored circles, and they are overlaid on the Neural Network predictions for the background, demonstrating the excellent agreement between simulation and fit. The points visually merge with the surface, indicating the high precision of the trained models, which we have already shown in Fig.~\eqref{fig:pressure_neuralnetwork}.

Additionally, the phase transition boundary to uniform nuclear matter, taken from \cite{ISHIZUKA2003517}, is superimposed on the same figure to delineate the region of validity for our approach. At very low temperatures ($T \approx 0$), the dominant degrees of freedom in this density regime are fully ionized nuclei and a degenerate electron gas. As temperature increases, thermally unbound neutrons emerge from the nuclei, forming a dilute neutron gas. However, this component does not dominate the thermodynamic quantities until the onset of the crust-core phase transition, as shown by the solid and dashed curves in Fig.~\eqref{fig:colorfig} and discussed in works such as \cite{ISHIZUKA2003517, dd2}.

Below the solid curve, the ionic component remains thermodynamically significant. Within the shaded uncertainty band of approximately 1~MeV width, model dependence is accounted for. In this region, the neutron gas is expected to contribute less than $\sim 30\%$ to the total baryonic content at the densities relevant to this study \citep{Grams2024}. Thus, our results remain applicable up to the transition to uniform matter, with the ionic sector continuing to play a dominant role in determining the thermodynamic response of the system.

\section{Thermal effect of ions in the outer Crust Eos}

\begin{figure}
    \centering
    \includegraphics[width=\linewidth]{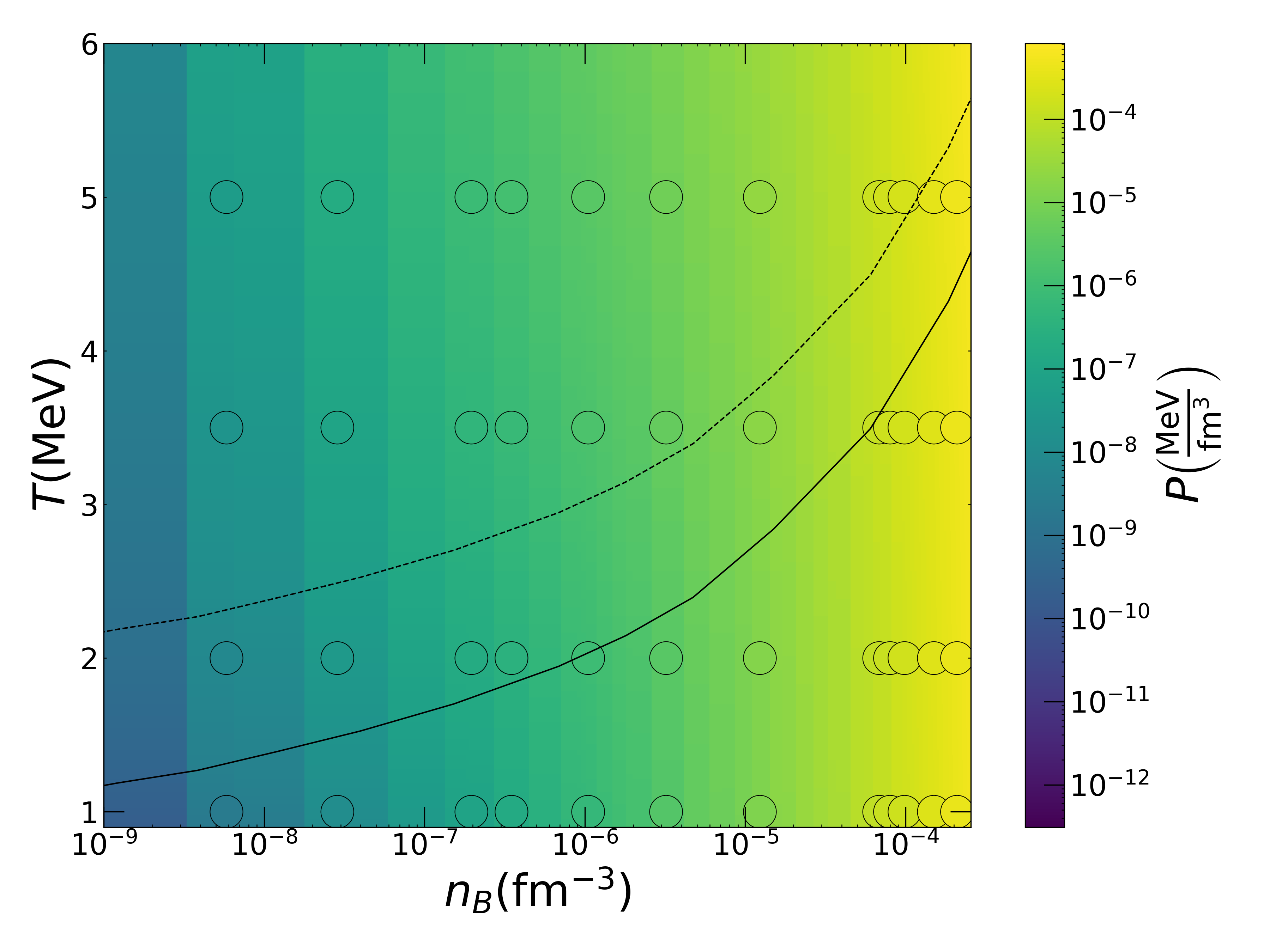}
    \caption{Pressure (color axis) as a function of baryonic density and temperature from our calculations, including ionic pressure from MD from the Neural Network parametrization and electron pressure from the degenerate gas. Individual points are also depicted coming directly from MD particular runs. We  show the region where the phase transition to uniform nuclear matter occurs between black dashed and solid lines, as taken from \cite{ISHIZUKA2003517}.}
    \label{fig:colorfig}
\end{figure}

\begin{figure}
    \centering
    \includegraphics[width=\linewidth]{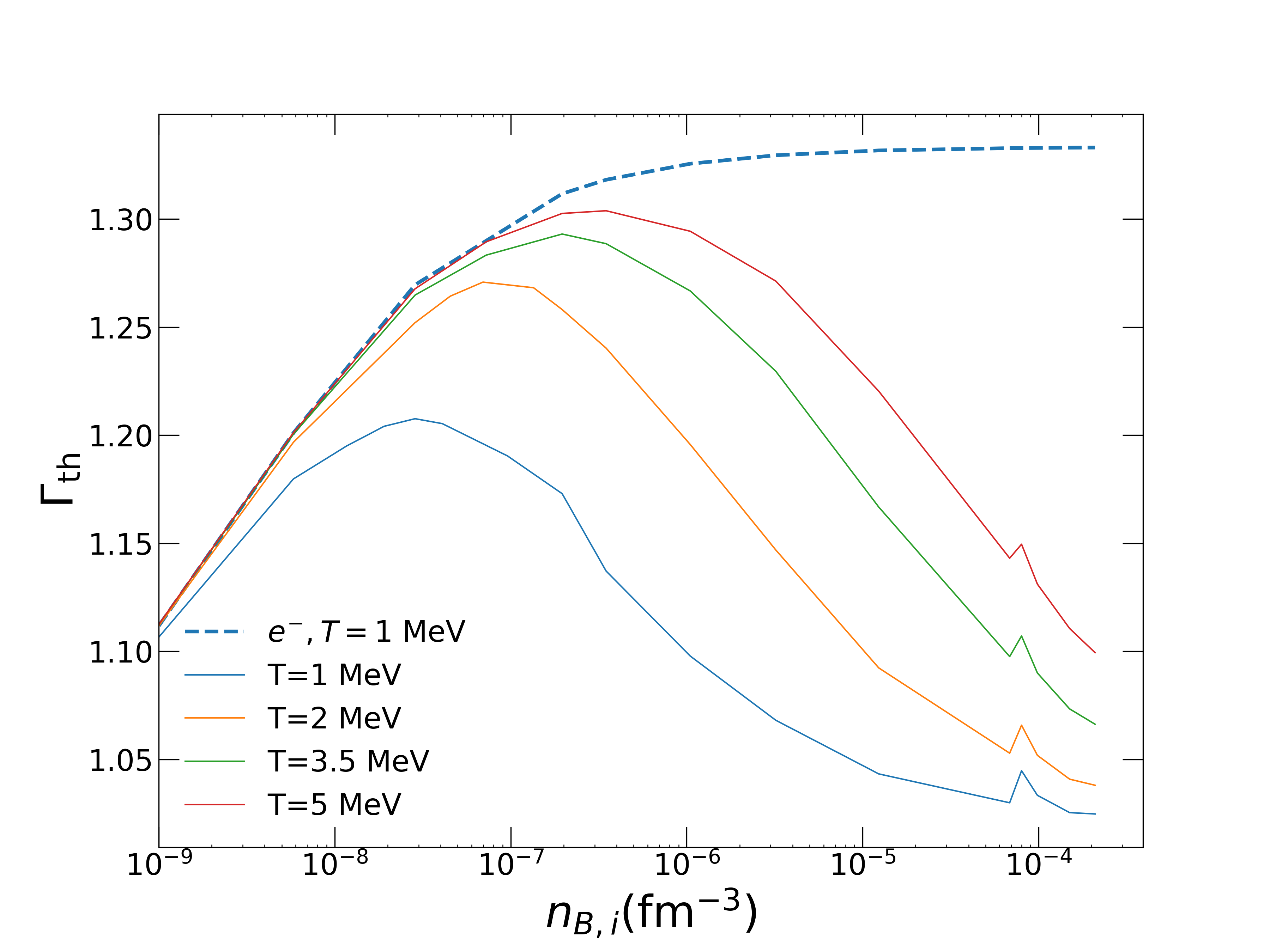}
    \caption{Thermal index $\Gamma_{th}$ as a function of baryonic density for several selected temperatures, as coming directly from MD. The simulated data points are tabulated in Table~\ref{tab:gammath}. We include (blue dashed line) the same $\Gamma_{th}$ for an electron gas, to showcase ionic effects.}
    \label{fig:gammath}
\end{figure}

We use the thermal index in Eq. \eqref{gamamth} to analyze the thermal effects of ions in the outer NS crust. Within the ion sector of the EoS we use for the zero temperature limit the ionic energy and pressure from \cite{Murarka_2022}

\begin{equation}
    \epsilon_i(T=0) = -C_1 Z^{5/3} k_{{F,e}} ; \quad 
        P_i(T=0) = -\frac{n_{B,i}}{3} C_1 \frac{Z^{5/3}}{A} k_{{F,e}}.
        \label{eq:ionszero}
\end{equation}
Here $C_1$ is the Madelung constant, which for the ground state of the ionic system, i.e. a \textit{bcc} lattice \citep{prc22} is $C_1=3.40665\times10^{-3}$ \citep{2008PhRvC..78b5807R}.

We calculate the thermal energy density and pressure from Eqs. \eqref{eq:eth} and \eqref{eq:pth}. For ions the finite-temperature quantities come from MD, while the zero temperature counterparts are those of Eq. \eqref{eq:ionszero}. In turn, electron thermal contributions are obtained from Eq. \eqref{eq:Pelec}.

In Fig.~\eqref{fig:gammath} we show $\Gamma_{th}$ as a function of baryonic density for our set of temperatures. Solid lines include both ions and electrons in the mixture, while dashed line depicts the $\Gamma_{th}$ only for an electron gas. There are several remarks to be made about its behavior, which shows a peak at intermediate crustal densities below the value expected for a relativistic gas, $\Gamma_{th}=4/3$. Thermal effects in the $P_{th}$, $\epsilon_{th}$ in our setting come from two different sources: the electron gas (Sommerfeld expansion) and ionic effects in the MD simulations. The simulated data points shown in this figure are tabulated in Table~\ref{tab:gammath}. The thermal part of the electron pressure, as obtained from Eq. \eqref{eq:Pelec} drops faster than its energy counterpart as density decreases, thus producing the tendency towards unity at low densities. However, closer to the inner crust boundary, where the electron-associated $\Gamma_{th}$ is expected to reach and maintain a $4/3$ value due to thermal effects being less relevant (see dashed line), the electron-ion combined $\Gamma_{th}$ undergoes a drop that is exclusively due to ionic thermal effects. This can be clearly seen in Fig.~\eqref{fig:gammath} by comparing solid and dashed blue lines, which only differ by the addition of ions in the latter. As at higher densities over the neutron drip the neutrons would start taking over the thermal effects, we show here the existence of a drop in $\Gamma_{th}$ below the relativistic e$^-$ gas value in a transitional region from the peak in Fig. \eqref{fig:gammath} up until the neutron drip density. This effect is solely due to the ionic thermal effects. We also note that the inherited composition variations at given baryonic densities have also an effect here seen at $n_{B,i}\sim 10^{-4} \ \rm fm^{-3}$ where the heavier nuclei appear in the sample. The small features visible appear as the ionic contribution to the energy density
develops a pronounced  sharp
increase and  change in slope, partly driven by terms scaling as
$Z^2(n_{B,i}/A)^{1/3}$, whereas the corresponding pressure contribution
shows a different density dependence, approximately
$Z^2(n_{B,i}/A)^{4/3}$. Since the thermal index is defined from the ratio 
$P_{\rm th}/\epsilon_{\rm th}$, these small differences
between the thermal pressure and energy density are enhanced. Similar composition induced features are also visible in Fig.~\eqref{fig:pressure_neuralnetwork} around that density and even at 
lower $n_{B,i}\sim 10^{-6} \, \rm fm^{-3}$ that is hardly visible, however. 
\begin{figure*}
    \centering
    \includegraphics[width=\linewidth]{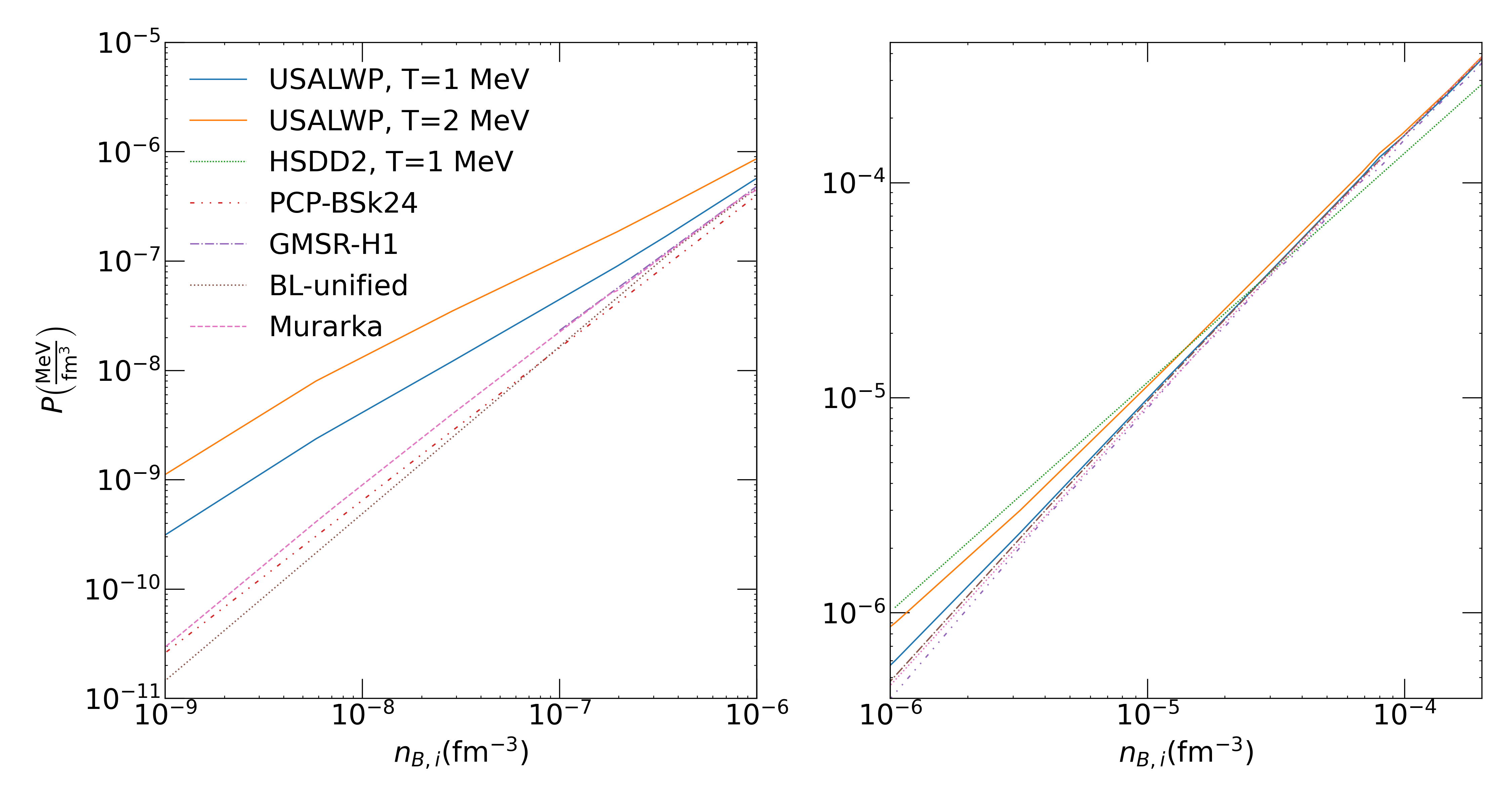}
    \caption{Pressure versus baryonic density for several equations of state in the literature, together with our data (USALWP) for $T=1,2$ MeV. We include equations of state DD2  at $T=1$ MeV \citep{eos_HSDD2} (green), and cold  PCP-BSk24 \citep{eos_pcpbsk24} (red), GMSR-H1 \citep{eos_GMSR} (purple), BL-unified \citep{eos_BL} (brown); and in pink the EoS whose composition we have utilized \citep{Murarka_2022}. Left and right panels depict different ranges of density for the sake of clarity. The $n_{B,i}=1\times 10^{-6} \rm fm^{-3}$ limit chosen indicates the lower confidence limit for the appearance of homogeneous matter at $T=2 \rm \ MeV$, see solid black line in Fig.~\eqref{fig:colorfig}.}
    \label{fig:eoscomparison}
\end{figure*}

Finally, we directly compare our work, combining MD results for ions with the Sommerfeld expansion for electrons, with several existing  equations of state in the crustal region in Fig.~\eqref{fig:eoscomparison}. Besides our model, USALWP, the models for the crust appearing in the figure, which we take from the CompOSE database \citep{Typel2015} are varied. HSDD2 \citep{eos_HSDD2}, GMSR-H1 \citep{eos_GMSR} and BL-unified \citep{eos_BL} use compressible liquid-drop models, the latter within a crust meta-model where parameters are extracted from a Bayesian analysis. In the case of PCP-BSk24 \citep{eos_pcpbsk24} the EoS is constructed within a unified crust-core framework based on energy-density functionals. 

This figure  illustrates the importance of treating thermal
effects consistently in the low-density crust. The four lower curves
(PCP-BSk24, GMSR-H1, BL-unified, and Murarka) correspond to published
zero-temperature EoSs, whereas HSDD2 and our USALWP results include
finite-temperature effects. In the low-density region shown in the left
panel, the finite-temperature curves largely depart from the cold
ones. In our calculation this enhancement results from the combined
thermal response of the ionic and electronic sectors i.e. the ions are
described including finite-size effects, screened interactions, and
correlations, while the thermal correction to the degenerate electron
gas is included through the Sommerfeld expansion as described before. Therefore, the spread between the warm and cold EoSs in this regime should not be interpreted simply as an uncertainty among equivalent models, since they incorporate
different physical ingredients. 
As the density increases, the curves progressively approach each other, as
shown in the right panel, reflecting the increasing degeneracy of the system
and the reduced importance of thermal corrections. We have chosen such temperatures, $T=1,2$ MeV,  as they lie fully within the mixed phase region as seen in Fig. \eqref{fig:colorfig}. Therefore, for applications involving a warm, low-density crust, an EoS including ionic correlations and thermal electron effects captures a consistent description rather than a strictly
zero-temperature  one whereas these differences become much smaller
toward neutron drip densities. As a result the thermal effects are seen as most important at the lowest densities where the Coulomb parameter $\Gamma_C\approx 0.3$ and thus thermal effects are more relevant.
Usual approaches to finite temperature extensions of these works usually assume an ideal classical gas of baryons plus a small fraction of light nuclei even at temperatures and densities within the mixed-phase region of the phase diagram.

\section{Conclusions}
We have performed MD simulations at densities 
corresponding to the outer crust of a Neutron Star, at finite 
temperature. Using the efficient Ewald summation and combining a 
screening electron gas with  finite-sized ions we compute the 
EoS of low-density matter as a function of baryonic 
density $n_B$ and temperature $T$. We assume that the cold
composition does not alter much with thermal energies of a few MeV.
The energy density and pressure in the crust are computed from data obtained from thermally equilibrated samples 
incorporating short-, long-distance effects from the
interactions, that exhibit screening behavior due to electrons parametrized  through the 
Thomas-Fermi wave vector. We have incorporated electrons 
to the EoS via a relativistic free gas at
finite temperature using the Sommerfeld expansion.

We have produced a Neural Network parametrization with the
aim of accurately predicting intermediate values of density
and temperature not directly computing by our simulations,
considering the same composition at each density range from
previous calculations in the literature.

Our calculation shows the impact of finite-temperature effects in the outer
crust, which we have analyzed and compared with a set of zero and thermal  calculations.
In particular, our EoS treatment of the warm crust consistently includes ionic correlations, screened finite-size interactions, and thermal corrections in the electron sector. We have calculated
the $\Gamma_{\rm th}$ factor and found it to be quenched with respect to values obtained from other treatments in the literature. Ionic thermal effects become particularly relevant at higher densities, producing a decrease of $\Gamma_{\rm th}$ toward the inner-crust boundary. These results indicate that the additional thermal and correlation effects included in our approach can be important for an accurate description of the warm outer crust.

\section{DATA AVAILABILITY}

The data and scripts underlying this article are available in Zenodo at https://doi.org/10.5281/zenodo.15348712. 

\section*{Acknowledgments}
D.B.G., C.A. and M.A.P.G. acknowledge partial support from the Spanish Ministry of Science PID2025-168032NB-I00, PID2022137887NB-100, RED2022-134411-T, Junta de Castilla y Le\'on SA101P24, SA091P24 and RES resources under AECT-2023-1-0026, AECT-2024-2-0009 projects. D.B.G. acknowledges support from a previous Ph.D. Fellowship funded by Programa Operativo and Consejería de Educación de la Junta de Castilla y León and European Social Fund Plus.

\bibliography{biblio}{}
\bibliographystyle{aasjournal}

\appendix

\begin{table*}[h!]
\centering
\begin{tabular}{ccc|ccc||ccc}
\hline $T \left(\rm MeV\right)$ & $\epsilon \left(\frac{\rm MeV}{\rm fm^3}\right)$ & $P_i \left(\frac{\rm MeV}{\rm fm^3}\right)$ & $T \left(\rm MeV\right)$ & $\epsilon \left(\frac{\rm MeV}{\rm fm^3}\right)$ & $P_i \left(\frac{\rm MeV}{\rm fm^3}\right)$ & Z & A & $n_{B,i} \left(\rm fm^{-3}\right)$ \\
1 & $1.9005\times 10^{-11}$ & $1.3022\times 10^{-11}$ & 2 & $3.9410\times 10^{-11}$ & $2.6348\times 10^{-11}$ & $26$ & $56$ & $7.4878\times 10^{-10}$\\
1 & $1.1747\times 10^{-10}$ & $8.6777\times 10^{-11}$ & 2 & $2.6416\times 10^{-10}$ & $1.8100\times 10^{-10}$ & $28$ & $62$ & $5.8253\times 10^{-9}$\\
1 & $4.9702\times 10^{-10}$ & $4.7169\times 10^{-10}$ & 2 & $1.2983\times 10^{-9}$ & $9.5748\times 10^{-10}$ & $28$ & $62$ & $2.8613\times 10^{-8}$\\
1 & $1.7086\times 10^{-9}$ & $3.0410\times 10^{-9}$ & 2 & $6.8923\times 10^{-9}$ & $6.3570\times 10^{-9}$ & $28$ & $64$ & $1.9606\times 10^{-7}$\\
1 & $-5.3850\times 10^{-11}$ & $1.6799\times 10^{-9}$ & 2 & $1.0109\times 10^{-8}$ & $1.0105\times 10^{-8}$ & $28$ & $64$ & $3.4870\times 10^{-7}$\\
1 & $-2.1513\times 10^{-8}$ & $3.1218\times 10^{-9}$ & 2 & $1.2392\times 10^{-9}$ & $1.7255\times 10^{-8}$ & $36$ & $86$ & $1.0486\times 10^{-6}$\\
1 & $-1.0046\times 10^{-7}$ & $1.3256\times 10^{-8}$ & 2 & $-2.6980\times 10^{-8}$ & $5.8404\times 10^{-8}$ & $34$ & $84$ & $3.2106\times 10^{-6}$\\
1 & $-6.8590\times 10^{-7}$ & $5.6693\times 10^{-9}$ & 2 & $-3.8268\times 10^{-7}$ & $1.8951\times 10^{-7}$ & $32$ & $82$ & $1.2342\times 10^{-5}$\\
1 & $-5.8492\times 10^{-6}$ & $6.6980\times 10^{-8}$ & 2 & $-4.0402\times 10^{-6}$ & $1.1415\times 10^{-6}$ & $28$ & $80$ & $6.8476\times 10^{-5}$\\
1 & $-7.2882\times 10^{-6}$ & $2.8953\times 10^{-8}$ & 2 & $-5.2009\times 10^{-6}$ & $1.2697\times 10^{-6}$ & $28$ & $80$ & $8.0000\times 10^{-5}$\\
1 & $-1.2941\times 10^{-5}$ & $-7.7624\times 10^{-7}$ & 2 & $-1.1065\times 10^{-5}$ & $3.1370\times 10^{-7}$ & $41$ & $122$ & $9.8448\times 10^{-5}$\\
1 & $-2.1393\times 10^{-5}$ & $-3.6458\times 10^{-6}$ & 2 & $-1.8435\times 10^{-5}$ & $-1.9393\times 10^{-6}$ & $38$ & $119$ & $1.5009\times 10^{-4}$\\
1 & $-3.0911\times 10^{-5}$ & $-4.7140\times 10^{-6}$ & 2 & $-2.6690\times 10^{-5}$ & $-2.2732\times 10^{-6}$ & $36$ & $116$ & $2.0938\times 10^{-4}$\\
\hline \hline
3.5 & $6.9802\times 10^{-11}$ & $4.6522\times 10^{-11}$ & 5 & $9.9973\times 10^{-11}$ & $6.6624\times 10^{-11}$ & $26$ & $56$ & $7.4879\times 10^{-10}$\\
3.5 & $4.7961\times 10^{-10}$ & $3.2308\times 10^{-10}$ & 5 & $6.9498\times 10^{-10}$ & $4.6521\times 10^{-10}$ & $28$ & $62$ & $5.8253\times 10^{-9}$\\
3.5 & $2.1615\times 10^{-9}$ & $1.6688\times 10^{-9}$ & 5 & $3.3875\times 10^{-9}$ & $2.3721\times 10^{-9}$ & $28$ & $62$ & $2.8613\times 10^{-8}$\\
3.5 & $1.4316\times 10^{-8}$ & $1.1193\times 10^{-8}$ & 5 & $2.3440\times 10^{-8}$ & $1.5170\times 10^{-8}$ & $28$ & $64$ & $1.9606\times 10^{-7}$\\
3.5 & $2.3506\times 10^{-8}$ & $1.8779\times 10^{-8}$ & 5 & $3.6517\times 10^{-8}$ & $2.7282\times 10^{-8}$ & $28$ & $64$ & $3.4870\times 10^{-7}$\\
3.5 & $3.3084\times 10^{-8}$ & $3.7444\times 10^{-8}$ & 5 & $6.3827\times 10^{-8}$ & $5.7210\times 10^{-8}$ & $36$ & $86$ & $1.0486\times 10^{-6}$\\
3.5 & $7.6622\times 10^{-8}$ & $1.2331\times 10^{-7}$ & 5 & $1.7473\times 10^{-7}$ & $1.8596\times 10^{-7}$ & $34$ & $84$ & $3.2106\times 10^{-6}$\\
3.5 & $3.9841\times 10^{-8}$ & $4.5066\times 10^{-7}$ & 5 & $4.4317\times 10^{-7}$ & $7.0359\times 10^{-7}$ & $32$ & $82$ & $1.2342\times 10^{-5}$\\
3.5 & $-1.5609\times 10^{-6}$ & $2.6551\times 10^{-6}$ & 5 & $8.0200\times 10^{-7}$ & $4.1223\times 10^{-6}$ & $28$ & $80$ & $6.8476\times 10^{-5}$\\
3.5 & $-2.2609\times 10^{-6}$ & $3.0567\times 10^{-6}$ & 5 & $4.9677\times 10^{-7}$ & $4.7710\times 10^{-6}$ & $28$ & $80$ & $8.0000\times 10^{-5}$\\
3.5 & $-8.4792\times 10^{-6}$ & $1.8600\times 10^{-6}$ & 5 & $-6.0580\times 10^{-6}$ & $3.3330\times 10^{-6}$ & $41$ & $122$ & $9.8448\times 10^{-5}$\\
3.5 & $-1.4422\times 10^{-5}$ & $4.5717\times 10^{-7}$ & 5 & $-1.0376\times 10^{-5}$ & $2.9333\times 10^{-6}$ & $38$ & $119$ & $1.5010\times 10^{-4}$\\
3.5 & $-2.0896\times 10^{-5}$ & $1.1702\times 10^{-6}$ & 5 & $-1.5520\times 10^{-5}$ & $4.4656\times 10^{-6}$ & $36$ & $116$ & $2.0938\times 10^{-4}$\\
\end{tabular}
\caption{Molecular Dynamics results for energy density and pressure for runs with different temperatures, for densities and composition in the Neutron Star (NS) outer crust assumed in \cite{Murarka_2022} with a representative heavy nucleus for each density in the absence of neutron gas.}
\label{tab:md_outer_crust}
\end{table*}
\begin{table*}[h!]
\centering
\label{tab:gammath}
\begin{tabular}{lrrrr}
\toprule
$n_{B,i} \left(\rm fm^{-3}\right)$ & $\Gamma_{\rm th}(T=1)$ & $\Gamma_{\rm th}(T=2)$ & $\Gamma_{\rm th}(T=3.5)$ & $\Gamma_{\rm th}(T=5)$ \\
$7.4878\times 10^{-10}$ & 1.095 & 1.098 & 1.098 & 1.098 \\
$5.8253\times 10^{-9}$ & 1.180 & 1.197 & 1.200 & 1.201 \\
$2.8613\times 10^{-8}$ & 1.208 & 1.252 & 1.265 & 1.268 \\
$1.9606\times 10^{-7}$ & 1.173 & 1.258 & 1.293 & 1.303 \\
$3.4870\times 10^{-7}$ & 1.137 & 1.240 & 1.289 & 1.304 \\
$1.0486\times 10^{-6}$ & 1.098 & 1.196 & 1.267 & 1.294 \\
$3.2106\times 10^{-6}$ & 1.068 & 1.147 & 1.230 & 1.271 \\
$1.2342\times 10^{-5}$ & 1.043 & 1.092 & 1.167 & 1.220 \\
$6.8476\times 10^{-5}$ & 1.030 & 1.053 & 1.098 & 1.143 \\
$8.0000\times 10^{-5}$ & 1.045 & 1.066 & 1.107 & 1.150 \\
$9.8448\times 10^{-5}$ & 1.033 & 1.052 & 1.090 & 1.131 \\
$1.5009\times 10^{-4}$ & 1.025 & 1.041 & 1.073 & 1.111 \\
$2.0938\times 10^{-4}$ & 1.025 & 1.038 & 1.066 & 1.099 \\
\end{tabular}
\caption{$\Gamma_{\rm th}$ for our EoS as a function of $n_{B,i}$ for different temperatures (in MeV), as shown in Fig.~\eqref{fig:gammath}.}
\end{table*}


\end{document}